\documentclass[structabstract]{aa}  
\usepackage{graphicx}
\usepackage{txfonts}
\begin{document}
   \title{Prominent spiral arms in the gaseous outer galaxy disks}


   \author{G. Bertin
          \inst{1}
          \and
          N.C. Amorisco\inst{2}
          }

   \institute{Dipartimento di Fisica, Universit\`a degli Studi di Milano,
via Celoria 16, I-20133 Milano, Italy\\
              \email{giuseppe.bertin@unimi.it}
         \and
             Dipartimento di Fisica, Universit\`a di Pisa, Largo Bruno
Pontecorvo 3, I-56127 Pisa, Italy\\
             \email{n.amorisco@sns.it}
             \thanks{also at Scuola Normale Superiore, Piazza dei
Cavalieri 7, I-56125 Pisa, Italy; as of October 2009, at Institute
of Astronomy, Madingley Road, Cambridge CB3 0HA, UK}
             }

   \date{Received November, 2009; }

 
  \abstract
   {Several spiral galaxies, as beautifully exhibited by the case of
NGC 6946, display a prominent large-scale spiral structure in
their gaseous outer disk. Such structure is often thought to pose
a dynamical puzzle, because grand-design spiral structure is
traditionally interpreted as the result of density waves carried
mostly in the stellar disk.}
   {Here we argue that the outer spiral
arms in the cold gas outside the bright optical disk actually have a
natural interpretation as the manifestation of the mechanism that
excites grand-design spiral structure in the main, star-dominated
body of the disk: the excitation is driven by angular momentum
transport to the outer regions, through trailing density waves
outside the corotation circle that can penetrate beyond the Outer
Lindblad Resonance in the gaseous component of the disk.}
   {Because
of conservation of the density wave action, these outgoing waves
are likely to become more prominent in the outer disk and
eventually reach non-linear amplitudes. To calculate the desired amplitude profiles, we make use of the theory of dispersive waves.}
   {If the conditions beyond
the optical radius allow for an approximate treatment in terms of
a linear theory, we show that fitting the observed amplitude
profiles leads to a quantitative test on the density of the disk
material and thus on the dark matter distribution in the outer
parts of the galaxy.}
   {This study is thus of interest to the
general problem of the disk-halo decomposition of rotation curves.}


   \maketitle
%

\section{Introduction}

Deep HI observations of nearby galaxies have led to the discovery
of a number of important phenomena that are changing our views on
the structure and dynamics of galaxies. These include the
existence in early-type galaxies of regular and radially extended
HI disks (Oosterloo et al. 2007a), the presence in spiral galaxies
of extraplanar gas characterized by slow rotation (for NGC 891 see Oosterloo et al. 2007b; for 
NGC 2403, see Fraternali et al. 2002), and the properties of
small-scale structures in the HI distribution (for NGC 6946, see
Boomsma et al. 2008). One interesting related discovery is the
existence of regular and prominent spiral arms in the gaseous
outer disk, well outside the bright optical disk (Shostak \& van der
Kruit 1984; Dickey et al. 1990; Kamphuis 1993; for NGC 2915, see
Meurer et al. 1996; for NGC 3741, see Begum et al. 2005). In this
respect, a particularly impressive example is given by the case of
NGC 6946 (Boomsma, 2007; Boomsma et al. 2008; see Fig.~1), where a spectacular
set of gaseous arms can be traced all the way out, with a
significant degree of regularity and symmetry even if the outer
disk is clearly lopsided and characterized by a fragmentary
structure; the outer arms also appear to contain stars (see
Ferguson et al. 1998, who also analyze the interesting cases of
NGC 628 and NGC 1058, and Sancisi et al. 2008).

The study of spiral structure in galaxies has received great
attention in the past. It is now generally thought that
grand-design structure is the manifestation of density waves,
mostly carried by the stellar component of galaxy disks. A general
framework for the interpretation of the observed morphologies,
based on a density wave theory, has received significant support
from the observations of spiral galaxies in the near infrared (see
Bertin \& Lin 1996 and references therein). It might thus appear
as a puzzle to find well-organized spiral patterns in the outer
disk, in a region where stars are practically absent.

The above-mentioned deep HI studies also serve as interesting
probes in view of defining an appropriate visible matter - dark
halo decomposition of the gravitational field in galaxies. In
particular, the studies of prominent spiral arms in the gaseous
outer disks have raised two issues  that separately point to the
question of whether the outer disk is light or heavy. On the one
hand, concerns have been raised about the applicability, in such
region, of the criterion for the onset of star formation proposed
by Kennicutt (1989), which relies on a threshold on the
axisymmetric stability parameter $Q = c \kappa/\pi G \sigma$.
Accordingly, light disks, with low $\sigma$, should be unable to
make new stars; but in the outer parts the disk might be flared
and three-dimensional effects may change the picture significantly
(in this regard, see also the comments made by Ferguson et al.
1998). On the other hand, there is widespread belief (see Sancisi et al. 2008, p. 212) that light
disks should be unable to support spiral structure (see Toomre
1981; Athanassoula et al. 1987; criticism against this belief can
be found, e.g., in the article by Bertin et al. 1989a). 

In general, the arguments often put forward in favor of a
maximum-disk decomposition of the rotation curves of spiral
galaxies (starting with van Albada \& Sancisi 1986) still await a
decisive measurement to remove the remaining degeneracy that
characterizes such a decomposition. Some projects, such as the
``Disk Mass Project" (Verheijen et al. 2004, 2007), aim at making
full use of three-dimensional gas and stellar dynamical data so as
to decompose the field, much like in the classical problem of the
disk thickness in the solar neighborhood (Oort 1932, Bahcall 1984,
Kuijken \& Gilmore 1989, Cr\'ez\'e et al. 1998, Holmberg \& Flynn
2000, 2004). Dark halos are generally thought to be made of
collisionless dark matter and to have spheroidal shape, but it
would be important to have direct proof that the outer disk of spiral
galaxies is indeed light, in contrast with the possibility that
the disk be heavy because of large amounts of molecular material
(Pfenniger et al. 1994; see also Revaz et al. 2009).

One interesting aspect of galactic dynamics is that models for the
interpretation of observed structures generally offer an
independent diagnostics of the overall mass distribution. In this
paper we will present one more example of this general aspect of
dynamics. Indeed, we will propose a model for the observed spiral
structure in the outer disk and then will show how the model can
be tested and applied to probe the structure of the outer disk in
relation with the problem of the amount and distribution of dark
matter.

In passing, we note that, in principle, one might test the properties 
of the final model (i.e., disk-dark matter decomposition) identified 
by our technique in specific cases against the expectations of non-Newtonian 
theories of gravity such as MOND; but of course, a discussion within MOND of 
the full problem, including the behavior of density waves, is currently not available.

The picture presented in this paper is the following. Global
spiral modes are driven by the transfer of angular momentum to the
outer regions (see Bertin et al. 1989a,b and references therein;
see also Lynden-Bell \& Kalnajs 1972, Bertin 1983). Outside the
corotation circle, the transfer is performed by short trailing
waves. At the Outer Lindblad Resonance, such outgoing waves are
fully absorbed in the stellar disk (Mark 1971, 1974), but only
partially absorbed in the gaseous component (Pannatoni 1983; Haass
1983), so that the signal can penetrate beyond such resonance and
propagate in the HI disk, if present. The outer spiral arms are
thus interpreted as the natural extension in the outer disk of the
short trailing waves that are responsible for exciting the global
spiral structure in the star-dominated main body of the galaxy
disk. The amplitude profile of such outer arms should simply
conform to the requirements dictated by the conservation of wave
action (Shu 1970; Dewar 1972). Because of this conservation, the
amplitude of the outer arms is expected to increase with radius,
in the regions where the inertia of the disk becomes smaller and
smaller, much like ocean waves can reach high amplitudes when
moving close to the shore.  In these outer regions, the density
wave is thus carried by the gas, but the stars present would
collectively respond and some new stars may be born because of gas
compression, following the gaseous arms.

Of course, galaxies such as NGC 6946 or
NGC 628 and the Blue Compact Dwarf NGC 2915 are very different
objects; each case should thus be studied separately in detail and
each individual object may have its own special character. Here we
wish to offer one quantitative reference frame for a common mechanism that in
general should operate in the outer parts of galaxy disks.

A fully non-linear, three-dimensional analysis of the properties
of density waves in the HI outer disk is not available, but we can
hope that under suitable conditions an approximate description
based on the laws of conservation of wave action for low-amplitude
density waves is viable. In any case, it should be tested against
the observations. Since fitting the data requires assumptions on
the density associated with the spiral arms and the density
associated with the fluid basic state, a test of this scenario
would be able to tell how much mass in the outer disk is present
in the form of molecular gas.

   \begin{figure}
   \centering
\includegraphics[width=8cm]{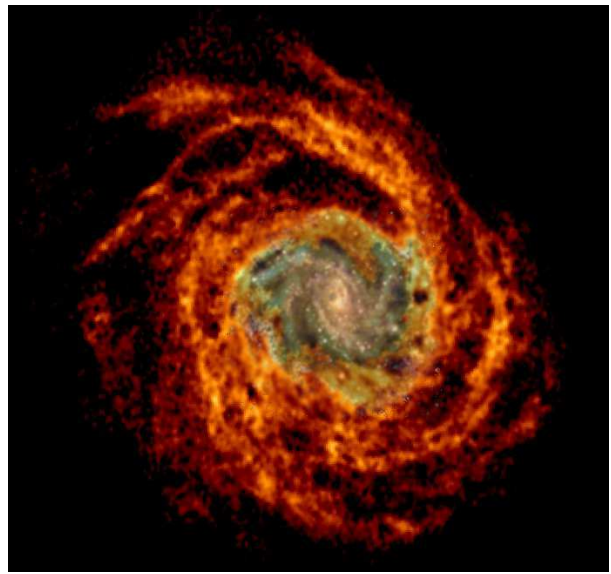}
      \caption{Total HI distribution of
NGC 6946 superimposed to its optical image; courtesy of Filippo
Fraternali (see Boomsma et al.~2008). 
              }
         \label{Fig1}
   \end{figure}

\section{Physical model and calculation of the amplitude profiles}

We refer to a grand-design spiral galaxy dominated by a global
mode with $m$ arms and pattern frequency $\Omega_p$ and, for
simplicity, consider a barotropic fluid model of an
infinitesimally thin disk as an idealized representation of the
gaseous outer regions of the galaxy. In terms of standard polar
cylindrical coordinates $(r, \theta)$, let $\sigma = \sigma_0 +
\sigma_1$ be the disk mass density, $(r \Omega +
v_1)\overrightarrow{e}_{\theta} + u_1 \overrightarrow{e}_r$ the
fluid velocity field, and $c$ the effective sound speed of the
fluid. The quantity $\Omega = \Omega (r)$ is the differential
rotation. As a measure of the distance from the corotation radius,
we then refer to the dimensionless quantity $\nu = m (\Omega_p -
\Omega)/\kappa$, where $\kappa$ is the epicyclic frequency. In the
following, subscript $0$ refers to the axisymmetric basic state of
the disk and subscript $1$ to the perturbation, on such
equilibrium, associated with the large-scale spiral structure. In
the notation just introduced, the relevant axisymmetric stability
parameter is defined as $Q = c \kappa/(\pi G \sigma_0)$.

According to the general picture of the modal theory of spiral
structure (see Bertin \& Lin 1996, and references therein),
outside the corotation circle the global mode is associated with
an outgoing (short) trailing wave. In general, for a normal spiral
mode the corotation circle is expected to occur at the edge of the
optical disk (i.e., at $r_{co} \approx 3 - 4 h$, with $h$ the
exponential scale of the stellar disk), while for a barred spiral
mode the corotation circle is expected to occur just outside the
tip of the bar, in the middle of the optical disk, at $r_{co}
\approx 2 h$. In the gaseous outer disk, the outgoing trailing
wave can penetrate beyond the Outer Lindblad Resonance (which
occurs at the radius $r_{OLR}$ where $\nu = 1$). This Section is
aimed at calculating the amplitude profile of the spiral arms in
the outer regions, for $r
> r_{OLR}$.  In this region the
calculation of the amplitude profile is particularly simple;
instead, the properties of spiral structure out to $r \approx
r_{OLR}$ are determined by the processes that govern the main body
of the disk (see Bertin \& Lin 1996).

In the linear theory of density waves, the calculation can be
performed in a straightforward manner by imposing the conservation
of the density of wave action (see Shu 1970; Dewar 1972; Bertin
1983). Mathematically, this is equivalent to carrying out the
analysis that leads to the standard algebraic dispersion relation
for density waves $D(\nu,|\hat{k}|) = 0$, with $D(\nu,|\hat{k}|)
= \nu^2 - 1 - Q^2 \hat{k}^2/4 + |\hat{k}|$ (where the radial wavenumber
$k = \hat{k} \kappa^2/(2 \pi G \sigma_0)$ is associated with the
pitch angle of the spiral arms, $\tan i = m/(rk)$), to the next
order in the WKB expansion (e.g., see Bertin 2000, Chapter 17.3;
see Eqs.~(17.2), (17.22), (17.28), (17.30), and (17.42)). 

In particular, the linear density wave analysis of a zero-thickness barotropic fluid disk is best carried out in terms of the enthalpy perturbation $h_1 = (c^2/\sigma_0)\sigma_1$. The WKB asymptotic analysis then leads to a Schroedinger-like equation of the form

\begin{equation}\label{Schro}
u'' + g(r,\omega)u = 0~,
\end{equation}

\noindent where, for tightly wound density waves, the function $g$ is given approximately by

\begin{equation}\label{effpot}
g(r,\omega) = \frac{\kappa^2}{c^2}\left( \frac{1}{Q^2} -1 + \nu^2\right)~.
\end{equation}

\noindent The function $u$ in Eq.~(\ref{Schro}) is defined in terms of the function $h_1$ in such a way that the amplitudes of the two functions are related as

\begin{equation}\label{effpot}
|h_1|^2 =  \frac{\kappa^2|1 - \nu^2|}{r\sigma_0}|u|^2 ~.
\end{equation}

\noindent The WKB analysis of the turning-point equation for $u$ shows that, away from the turning points, the amplitude of $u$ must scale as $|u| \sim |g|^{-1/4}$, which is interpreted as the conservation of the density of wave action. A little algebra then shows that such relation is equivalent to the condition

\begin{equation}\label{density}
\left|\frac{\sigma_1}{\sigma_0}\right|^2 \propto
G(\nu,Q) r^{-1}\kappa^{4}\sigma_0^{-4}~,
\end{equation}

\noindent where

\begin{equation}
G(\nu,Q) \equiv \frac{\nu^2 - 1}{Q^2 \sqrt{1 + (\nu^2 - 1)Q^2}}~.
\end{equation}

\noindent The proportionality constant is independent of $r$. Thus
Eq.~(\ref{density}) provides the desired expression for the
amplitude profile of the density wave in the outer regions.

In order to complete the description of the amplitude profiles of
the spiral arms, we may then consider the linearized continuity
equation:

\begin{equation}\label{cont}
\frac{\partial \sigma_1}{\partial t} + \frac{1}{r}
\frac{\partial}{\partial r} (r \sigma_0 u_1) + \frac{1}{r}
\frac{\partial}{\partial \theta} (\sigma_0 v_1 + \sigma_1 r
\Omega) = 0~.
\end{equation}

\noindent Since the vorticity equation shows that $v_1 = (i/\nu)
[\kappa/(2 \Omega)] u_1$, so that $v_1$ and $u_1$ have the same
order of magnitude, in the continuity equation we can neglect the
$v_1$ contribution consistent with the present WKB approximation,
thus obtaining:

\begin{equation}
\sigma_0 k u_1 \sim \nu \kappa \sigma_1~.
\end{equation}

\noindent By inserting here the solution for $k$ associated with
the (short) trailing wave branch,

\begin{equation}\label{shorttrailing}
k = - \left( \frac{\kappa^2}{2 \pi G \sigma_0}\right)
\frac{2}{Q^2}\left(1 + \sqrt{1+ (\nu^2 - 1)Q^2}\right)~,
\end{equation}

\noindent we find:

\begin{equation}\label{velocity}
\left|\frac{u_1}{r \Omega}\right| =\nu \left(\frac{\kappa}{\Omega}\right) \left(\frac{1}{|rk|}\right)\left|\frac{\sigma_1}{\sigma_0}\right|\propto  \left(
\frac{\kappa}{\Omega}\right) H(\nu,Q) r^{-3/2} \sigma_0^{-1}~,
\end{equation}

\noindent with

\begin{equation}
H(\nu,Q) \equiv \frac{\nu Q^2 \sqrt{G(\nu,Q)}}{1+ \sqrt{1 + (\nu^2
- 1)Q^2}}~.
\end{equation}

\noindent Again, the proportionality factor is independent of $r$. Note that the first part of Eq.~(\ref{velocity}) is an equality, not a proportionality relation. In other words, in the context of the linear WKB theory, the scale in velocity amplitude $u_1$ is uniquely determined by the scale in the amplitude of the density wave $\sigma_1$. Note also that the perturbation $u_1$, together
with $v_1$, will generate the ``wiggles" in the velocity field
that are characteristic of density waves (for M81, see Visser
1977); in the observed cases where the amplitudes are large, a
non-linear theory is required for a quantitative comparison with
the observations.

\subsection{Effects of the flaring of the outer disk}

The analysis described previously can be generalized to include
the effects of finite thickness of the disk. In the discussion of
the dynamics of self-gravitating disks, these effects are often
ignored, but may actually play an important role; in our case,
significant effects would be naturally expected if in the outer
parts the disk is flared, as often observed. Qualitatively, these
effects should become significant for short waves, when the
wavelength of the density wave becomes comparable to the thickness
of the disk; in terms of local stability, finite-thickness effects
are known to be stabilizing, because they effectively ``dilute"
the gravity field. Quantitatively, they are approximately
incorporated by the following dispersion relation (see Vandervoort
1970, Yue 1982):

\begin{equation}\label{ft}
D_{ft}(\nu,|\hat{k}|, \hat{z}_0) = \nu^2 - 1 - \frac{1}{4}Q^2
\hat{k}^2+\frac{|\hat{k}|}{1+ |\hat{k}|\hat{z}_0} = 0~,
\label{ftdispersionr}
\end{equation}

\noindent where $z_0$ represents the disk thickness (defined in
such a way that the disk surface density $\sigma_0$ is related to
the volume density $\rho_0$ on the equatorial plane by the
expression $\sigma_0 = 2 \rho_0 z_0$) and

\begin{equation}\label{thickness}
\hat{z}_0 \equiv \frac{\kappa^2 z_0}{2 \pi G \sigma_0}~.
\end{equation}

The modification of the dispersion relation with respect to the
standard one, used earlier in Sect.~2, changes the expression of
the trailing wave-branch that carries angular momentum outwards $k
= k_{ST} = \hat{k}_{ST} \kappa^2/(2 \pi G \sigma_0)$, with respect
to Eq.~(\ref{shorttrailing}). But the calculation of this
wavebranch is straightforward, because the new dispersion relation
(\ref{ft}) is just a cubic in $|k|$. Here $\hat{k}_{ST} =
\hat{k}_{ST} (\nu, Q, \hat{z}_0)$.

From the general theory of dispersive waves, we know that the flux
of density wave action is $\mathcal{F} = c_g \mathcal{A}$,
with the group velocity $c_g \equiv -(\partial \omega / \partial
k)$ and the wave action density $\mathcal{A} \propto \sigma_1^2
(\partial D/\partial \omega)$. Thus we have $\mathcal{F} \propto
(\partial D/ \partial k)$. In the derivation reported in the first
part of the Section the factor $(\partial D/ \partial k)$ enters
in the conservation equation (\ref{density}), as $(\partial D/
\partial \hat{k}) = \sqrt{1 +(\nu^2 - 1)Q^2}$, with the latter
derivative evaluated from the zero-thickness dispersion relation
on the short trailing wave-branch (\ref{shorttrailing}).

Therefore, the desired conservation equation in the finite
thickness case becomes:

\begin{equation}\label{densityft}
\left|\frac{\sigma_1}{\sigma_0}\right|^2 \propto G_{ft}(\nu,Q,
\hat{z}_0)r^{-1}\kappa^{4}\sigma_0^{-4}~,
\label{rapft}
\end{equation}

\noindent where now we have

\begin{equation}
G_{ft}(\nu,Q, \hat{z}_0) \equiv \frac{\nu^2 - 1}{Q^2 (\partial
D_{ft}/\partial \hat{k})}~;
\label{Gft}
\end{equation}

\noindent here the partial derivative is evaluated at $\hat{k} =
\hat{k}_{ST}$.

The corresponding expression for the amplitude profile in the
velocity field is obtained from Eq.~(\ref{velocity}) by replacing
$H(\nu, Q)$ with $H_{ft}(\nu, Q, \hat{z}_0)$, with

\begin{equation}
H_{ft}(\nu,Q, \hat{z}_0) \equiv \frac{\nu \sqrt{G_{ft}(\nu,Q,
\hat{z}_0)}}{|\hat{k}_{ST}(\nu,Q,\hat{z}_0)|}~.
\label{Hft}
\end{equation}

In the following subsection it will be shown that flaring effects
are under control and do not change significantly the general
predictions of the zero-thickness theory.

\subsection{A simple reference model}

To test the overall picture we may consider the following simple
reference model. We refer to a two-armed spiral structure ($m =
2$) in an outer disk characterized by a flat rotation curve, so
that $\kappa = \sqrt{2} \Omega$, $\Omega /\Omega_p = r_{co}/r$,
and $\nu = \sqrt{2}(r/r_{co} - 1)$. To be specific, we will assume
that $r_{co} = 3 h$, in terms of the exponential scale $h$ of the
(inner) stellar disk. Thus the Outer Lindblad Resonance will occur
at $r_{OLR} \approx 5.12 h$. Therefore, at $r_{in} = 6 h$ we are
outside the circle associated with the Outer Lindblad Resonance,
in a region where we expect the disk to be gas dominated, so that
beyond such radius we may proceed to apply our fluid model.

As a further simplification, we take the conservative case in
which $\sigma_0 = \sigma_0 (r_{in} )(6 h/r)$. This corresponds to
a very gentle decline of the gas density profile. (For faster
declining profiles, we expect that the relative strength of the
spiral amplitude should be more pronounced as we move outwards
(cf. Eq.~(\ref{density})), and thus give rise to a stronger
effect.) In our
simple reference model, the gas density is then proportional to
the differential rotation, so that we find $|\sigma_1/\sigma_0|^2
\propto G(\nu, Q)/r$ and $|u_1/(r\Omega)| \propto H(\nu,
Q)/r^{1/2}$.

In a zero-thickness disk, within the adopted reference model, the
pitch angle of the spiral arms depends only on $\nu$ and $Q$, because (cf.
Eq.~(\ref{shorttrailing}))

\begin{equation} \label{pitch}
|r k| \propto  \frac{1}{Q^2}\left(1 + \sqrt{1+(\nu^2 -
1)Q^2}\right)~,
\end{equation}

\noindent where the proportionality constant is independent of
radius. In this case we are thus left to discuss the dependence $Q
= Q(r)$. There are observational indications (cf. Boomsma et al.
2008, Fig.~6) that the gas velocity dispersion is steadily
declining in the outer disk, while in the present simple model the
quantity $\kappa/\sigma_0$ is $r$-independent. So we might be led
to argue that a {\it declining} $Q$-profile would be realistic.
However, the observed decline in the gas velocity dispersion is
just likely to reflect the fact that, because of the stabilizing
role of thickness (see previous subsection), the disk can get
colder and colder (in terms of the standard $Q$) and still remain
on the margin of local axisymmetric instability. Then, for the
present zero-thickness reference model we think it appropriate to
consider $Q \approx $ constant, and, for simplicity, we take $Q = 1$. Note
that this condition of marginal stability is not strictly
necessary, because in the outer disk short trailing waves can
propagate even at higher values of $Q$.

In conclusion, the present simple zero-thickness model is characterized by $G(\nu, Q) =
(\nu^2-1)/\nu$, $H(\nu, Q) = \sqrt{\nu(\nu^2 - 1)}/(1+\nu)$, and $rk
\propto (1 + \nu)$. By applying Eqs.~(\ref{shorttrailing}), (\ref{density}), and
(\ref{velocity}) we can proceed to
calculate the desired profiles $i(r)$, $|\sigma_1/\sigma_0|$, and
$|u_1/(r \Omega)|$.

In the finite thickness case, the wavenumber for short trailing
waves $k_{ST}(\nu, Q, \hat{z}_0)$ is obtained from the dispersion
relation Eq.~(\ref{ftdispersionr}), which can also be used to
calculate the two relevant quantities $G_{ft}(\nu, Q,\hat{z}_0)$
and $H_{ft}(\nu, Q,\hat{z}_0)$. The remaining point that requires
discussion is the radial dependence of the two functions
$\hat{z}_0=\hat{z}_0(r)$ and $Q = Q(r)$.

As to the vertical thickness of the outer disk, for our reference
model, characterized by $\Omega \sim 1/r$ and $\sigma_0 \sim 1/r$,
it can be shown (see Appendix A in Bertin \& Lodato 1999) that, in
each of the two opposite limits of a totally self-gravitating disk
and of a non-self-gravitating layer, the thickness behaves as $z_0
\sim r$, so that taking $\hat{z}_0=$constant is a reasonable
choice. In addition, since we are referring to a fluid, the
relevant velocity dispersion tensor is isotropic ($c_r = c_z =
c$), so that there is a one-to-one relation between the value of
$Q$ and the value of $\hat{z}_0$ that we are going to take; let us
denote this relation by $Q = Q_{fluid}(\hat{z}_0)$. On the other
hand, for a proper comparison with the zero-thickness reference
model described earlier in this Subsection, we should assume that
the disk is at marginal stability, i.e. $Q = Q_{max}(\hat{z}_0)$
(see Eq.~(15.22) in Bertin 2000). By combining the above
requirements into $Q_{fluid}(\hat{z}_0) = Q = Q_{max}(\hat{z}_0)$
we get a unique value for the pair $(\hat{z}_0, Q)$, as demonstrated in Fig.~2. With this
determination, from Eqs.~(\ref{ft}), (\ref{Gft}), (\ref{Hft}), and
(\ref{rapft}) we can calculate the desired profiles $i(r)$,
$|\sigma_1/\sigma_0|$, and $|u_1/(r \Omega)|$ for the finite
thickness model.

   \begin{figure}
   \centering
\includegraphics[width=8cm]{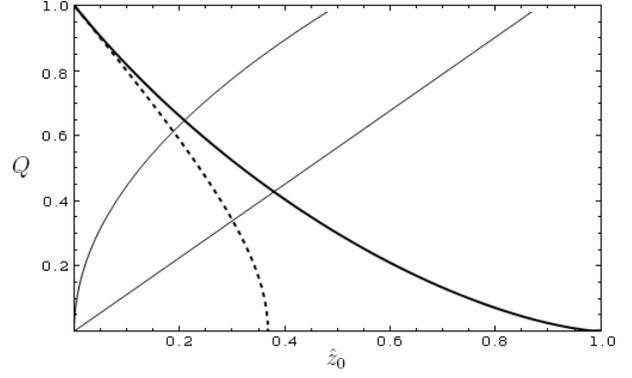}
      \caption{For our simple reference fluid model, with $\sigma_0 \propto 1/r$ and $\Omega \propto 1/r$, the thick solid line represents the marginal
              stability condition $Q = Q_{max}(\hat{z}_0)$ and the dashed curve the corresponding condition for the case in which the dilution of the gravity field (associated with the finite thickness of the disk) in the relevant dispersion relation is described by an exponential factor (instead of the rational factor used in Eq.~(\ref{ft})). The thin rising curves represent the function  $Q = Q_{fluid}(\hat{z}_0)$ for the two cases of a fully self-gravitating layer (upper curve) and for a non-self-gravitating layer (lower curve).}
         \label{Fig2}
   \end{figure}

The results of this analysis are illustrated in Fig.~3 in the
radial interval $6 h < r < 10 h$. In this figure, we have assumed
that at $r = 6 h$ the relative amplitude of the density wave is
$\sigma_1/\sigma_0 = 0.15$ and that the pitch angle of spiral
structure at such inner location is 15 degrees. The figure shows
that even for the presently assumed very gentle decline of the gas
density distribution the amplitude of the spiral wave steadily
increases with radius. Note that in the two limits, of a totally
self-gravitating disk and of a non-self-gravitating disk, the
finite-thickness effects do not change the general picture;
similar results and a similar conclusion have been checked to hold
using a dispersion relation alternative to
Eq.~(\ref{ftdispersionr}), in which the dilution of the
gravitational term $\left|\hat{k}\right|$ is exponential rather
than rational.

   \begin{figure}
   \centering
\includegraphics[width=8cm]{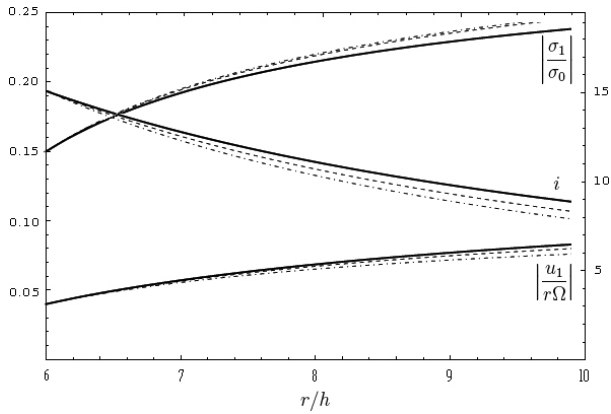}
      \caption{Relative amplitude
profiles of the spiral arms in
   terms of
    density $\sigma_1/\sigma_0$ and of radial velocity $u_1/(r\Omega)$
   (dimensionless, left axis) and pitch angle $i$ of spiral
    structure (degrees, right axis), for three cases of the simple reference model described in
Sect.~2.2: zero-thickness disk (thick lines), finite-thickness
totally self-gravitating disk (dashed lines), finite-thickness
non-self-gravitating disk (dash-dotted lines).
              }
         \label{Fig3}
   \end{figure}

\section{Discussion and conclusions}

HI observations of the gaseous outer regions provide measurements
of the HI gas density $\sigma_{HI}$, of the rotation curve
$\Omega$, and of the turbulent velocity $c_{HI}$, which we may
identify with $c$. We may argue that the actual disk density be
traced by the atomic hydrogen, so that $\sigma = f \sigma_{HI}$.
In the simplest model we may take $f \approx 1.4$, i.e. take that
the outer disk just contains the relevant proportion of helium.
For the following discussion, we may assume that the thickness
$z_0$ of the gaseous outer disk is not well constrained by the
observations, since we wish to consider objects with prominent
observed spiral arms, and therefore galaxies that are not edge-on.
On the other hand, instead of considering the thickness profile
$z_0 = z_0(r)$ as an additional free function of the problem, we
may refer to a self-consistent estimate of such profile, which is
readily available (e.g., see Appendix A in Bertin \& Lodato 1999
and Sect.~2.2 in the present paper).

Clearly, if the linear theory happens to be viable, the basic
relations are over-constrained by the data, since we only have
some leverage on the precise value of $\Omega_p$ (which then sets
the form of the function $\nu(r)$) and basically no other free
parameters. Note that the morphology, or a suitable Fourier
decomposition of the observed structure, would identify the
dominant value of $m$ and the pitch angle $i(r)$, i.e., the
function $k(r)$. The linear theory makes specific predictions,
such as Eqs.~(\ref{shorttrailing}), (\ref{density}),
(\ref{velocity}) (or the corresponding equations recorded in
Sect.~2.1 for the study that includes the effects of finite
thickness). Much as for the classical tests of the density wave
theory (for M81, see Visser 1977), we may hope that the various
observed quantities all fall within a reasonable agreement with
the theoretical expectations.

Suppose that we start from such a straightforward comparison with
the observations on the basis of the linear theory
described in the first part of Sect.~2. If this attempt turned out
to be unsatisfactory, we would have three levels of action in
interpreting the data.

(i) One possibility would be to make use of an HI fraction $f$
constant, but significantly larger than unity. Physically, a
choice of this kind would correspond to imagining a heavier outer
gaseous disk, but still in proportion to the observed HI disk, and
thus with little relevance to the overall problem of dark matter.
For given values of $\sigma_{HI}$ and $c_{HI}$, this would allow
us to reduce the value of the wavenumber scale $\kappa^2/(2 \pi G
\sigma_0)$ and of the axisymmetric stability parameter $Q$, while
leaving the relative density amplitude $\sigma_1/\sigma_0$
unchanged on the left-hand-side of Eq.~(\ref{density}). Except for
a small adjustment through the functions $G$ and $H$, such a
constant $f$ would have little or no effect on the fits to the
observed amplitude {\it profiles} dictated by Eqs.~(\ref{density})
and (\ref{velocity}).

(ii) A second possibility would be to make use of a free {\it
function} $f = f(r)$, with the general requirement that $f > 1.4$.
Resorting to such a function would correspond to imagining that
significant amounts of dark matter would be in the thin disk,
possibly in molecular form, with a distribution different from
that of the cold atomic hydrogen. This may thus go in the
direction of an alternative picture (with respect to the standard
picture of a spheroidal halo) for the general problem of dark
matter (e.g., see Pfenniger et al. 1994). Clearly a non-constant
HI fraction $f(r)$ would change the character of the observed
amplitude profiles predicted by Eqs.~(\ref{density}) and
(\ref{velocity}).

(iii) Finally, it may well be that the model developed above,
which relies on the predictions of a linear theory, turns out to
be inadequate. In other words, we should develop a model in which
the role of the non-linearities associated with the finite
amplitude observed in the prominent spiral arms is properly taken
into account.

While a more advanced model, of the kind outlined in item (iii)
above, is being developed (in this respect, we are encouraged by
the fact that a model of the outer disk as a fluid disk is likely
to be appropriate and that the physical picture considered, of an
outgoing wavetrain, is relatively simple), in this paper we argue
that a first test of the simple picture presented here is worth
trying. If a satisfactory agreement with the linear theory could
be obtained with $f \approx 1.4$, we would have one additional
convincing argument that the picture of a spheroidal dark halo
indeed holds also for the outermost regions of spiral galaxies. In
any case, the considerable effort required by setting up such a
test on a specific case, e.g. for NGC 6946, would prepare the
ground for a test of the more realistic non-linear theory that we plan to
investigate soon.

Finally, given the picture provided by the simple reference model
described in Sect.~2.2, we may argue that for those galaxies in
which the outer gaseous disk density declines too sharply, the
non-linear effects that would rapidly take place because of the
sharp increase in relative amplitude of the density waves may
actually tend to break the grand design outer structure and result
in turbulent dissipation, as often argued in the physical
discussion of the outer boundary condition for the establishment
of global spiral modes (e.g., see Bertin \& Lin 1996, p.~222).

\begin{acknowledgements}
     We wish to thank Renzo Sancisi, Rosemary Wyse, and Jay Gallagher
for pointing out the interest in this problem and for several
stimulating discussions and Filippo Fraternali and Giuseppe Lodato
for a number of useful comments. Special thanks to Filippo
Fraternali for providing us with Fig.~1.
\end{acknowledgements}

\end{document}